\documentstyle[prl,aps,preprint,tighten,floats,aps,epsf,psfig,axodraw]{revtex}

\def\be{\begin{equation}}
\def\ee{\end{equation}}
\def\bear{\begin{eqnarray}}
\def\eear{\end{eqnarray}}
\def\beqn{\begin{eqnarray}}
\def\eeqn{\end{eqnarray}}

\def\beq{\begin{equation} }
\def\eeq{\end{equation} }

\def\ben{\begin{eqnarray} }
\def\een{\end{eqnarray} }

\def\mod#1{{\rm (mod~2)} }

\def\ln{{\rm ln}\,}

\begin{document}
\draft
\date{\today}
\preprint{\vbox{\baselineskip=12pt
\rightline{UPR-xxx-T}
\vskip0.2truecm
\rightline{UM-TH-98/xx}
\vskip0.2truecm
\rightline{hep-ph/9902247}}}

\title{Physics Implications of a Perturbative Superstring 
Construction}
\author{M. Cveti\v c${}^{\dagger}$, L. Everett${}^{\dagger}{}^{*}$, P.
Langacker${}^{\dagger}$, and J. Wang${}^{\dagger}$}
\address{${}^{\dagger}$Department of Physics and Astronomy \\
          University of Pennsylvania, Philadelphia PA 19104-6396, USA \\
${}^{*}$Randall Laboratory of Physics, University of Michigan\\
Ann Arbor, MI 48109, USA}
\maketitle
\begin{abstract}
We investigate the low energy physics implications of a prototype
quasi-realistic superstring model with an anomalous $U(1)$.  First, we
present the techniques utilized to compute the mass spectrum and
superpotential couplings at the string scale, and demonstrate the results
for the effective theory along a particular flat direction/``restabilized
vacuum" of the model. We then analyze the gauge symmetry breaking
patterns and renormalization group equations to determine the mass
spectrum at the electroweak scale for a particular numerical example with 
a realistic $Z-Z'$ hierarchy. Although the model considered is not
fully realistic, the results demonstrate general features of
quasi-realistic string models, such as extra matter (e.g. $Z'$
gauge bosons and an extended Higgs sector) at the electroweak/TeV scale, 
and noncanonical couplings (such as $R$-parity violating terms).

\end{abstract}
%\vskip 0.3truecm
%\pacs{\tt PACS number(s): }
\vskip2cm
%\leftline{}
%\leftline{CERN-TH/97-338}
%\pacs{}

\section{Introduction}

Predictions from superstring theory provide natural possible extensions of
the MSSM.
However, there are several problems to be resolved in attempting to
connect string theory to the observable, low energy world. 
First, many models can be derived from string theory, and there is no
dynamical principle as yet to select among them. Furthermore, no fully
realistic model, i.e., a model which contains just the particle content
and couplings of the MSSM, has been constructed.  In addition, there is no
compelling scenario for how to break supersymmetry in string theory, and
so soft supersymmetry breaking parameters must be introduced into the
model by hand, just as in the MSSM.  

We adopt a more modest strategy and consider a class of quasi-realistic
models constructed within weakly coupled heterotic string theory.  In
addition to the necessary ingredients of the MSSM, such models generically
contain an extended gauge structure that includes a number of $U(1)$ gauge
groups and ``hidden'' sector non-Abelian groups, and 
additional matter fields (including a number of SM exotics and SM
singlets). They predict gauge couplings unification (sometimes with
non-standard Ka\v c-Moody level) at the string scale
$M_{String} \sim 5 \times 10^{17}$ GeV. 
%Although the presence of extra
%matter is potentially problematic, an important mechanism exists that can
%in principle decouple a number of the exotic fields at the string scale,
%as will be discussed in the subsequent section. 
The most desirable feature of models in this class is that the
superpotential is explicitly calculable; in particular, the non-zero Yukawa
couplings are ${\cal O}(1)$, and can naturally accommodate the radiative
electroweak symmetry breaking scenario.  In addition, string selection
rules can forbid gauge-allowed terms, in contrast to the case in general
field-theoretic models.  

In addressing the phenomenology of these models, there are two
complementary approaches. The first is the "bottom-up" approach, in which
models with particle content and couplings motivated from
quasi-realistic string models are studied to provide insight into the
new physics that can emerge from string theory (such as additional $Z'$
gauge bosons)\cite{cl,SY,cdeel,cceel1,lw}.  In this work we adopt the
second (``top-down") approach, and analyze a prototype string
model (Model 5 of \cite{chl}) in detail.  The analysis of this class of
string models (done in collaboration with G. Cleaver and J. R.
Espinosa) proceeds in several stages, which will be briefly summarized
below and is documented in \cite{cceel2,cceel3,cceelw1,cceelw2}.  We then
focus on the main results: the determination of the mass spectrum and 
trilinear couplings at the string scale, the renormalization group
analysis, the low energy gauge symmetry breaking patterns and the 
mass spectrum of the model at the electroweak scale.

Our analysis shows that the prototype model is not fully realistic.  
In particular, many of the SM exotics remain massless in the low
energy theory.  However, we find there are other general features of the
model which have interesting phenomenological implications, including an
additional low-energy $U(1)'$ gauge group, $R$ parity violating couplings,
``mixed" effective $\mu$ terms, and extended chargino, neutralino, and
Higgs sectors (with patterns of mass spectra that differ substantially
from the case of the MSSM).  

In section II, we discuss the generation of the effective mass terms and
the trilinear couplings associated with the flat direction. In section
III, we present the effective couplings and the implications of the
effective theory along a particular flat direction as an illustrative
example. We conclude in section IV.   

\section{Flat Directions and Effective Couplings}

The model we have chosen as a prototype model to analyze is Model 5 of
\cite{chl}. Prior to vacuum restabilization, the model has the gauge
group
\begin{equation}
\{SU(3)_C\times SU(2)_L\}_{\rm obs}\times\{SU(4)_2\times SU(2)_2\}_{\rm hid}
\times U(1)_A\times U(1)^6,
\end{equation}
and a particle content that includes the following
chiral superfields in addition to the MSSM fields:
\begin{eqnarray}
&&6 (1,2,1,1) + (3,1,1,1) + (\bar{3},1,1,1) + \nonumber\\
&&4 (1,2,1,2) + 2 (1,1,4,1) + 10 (1,1,\bar{4},1) +\nonumber\\
&&8 (1,1,1,2) + 5 (1,1,4,2) + (1,1,\bar{4},2) +\nonumber\\
&& 8 (1,1,6,1) + 3 (1,1,1,3)+ 42 (1,1,1,1)\;\;,
\end{eqnarray}
where the representation under $(SU(3)_C,SU(2)_L,
SU(4)_2,SU(2)_2)$ is indicated. The SM hypercharge is determined as a
linear combination of the six non-anomalous $U(1)$'s.   

%An important feature present in all of the quasi-realistic models
%considered is that one of the Abelian gauge group factors is anomalous
%(the trace of the $U(1)_A$ charge is nonzero). In string theory, there is
%a well-known mechanism which cancels the anomalies, at the expense of
%generating a constant term of ${\cal O}(M_{String})$ to the $D$ term of the
%anomalous $U(1)$.  
As the first step of the analysis, we address the presence of the 
anomalous $U(1)_A$ generic to this class of models. 
%A standard
%anomaly cancellation mechanism is present in the effective theory, as the
%underlying string theory is anomaly free. However, 
The standard anomaly cancellation mechanism generates a nonzero
Fayet-Iliopoulos  (FI) term of ${\cal O}(M_{String})$ to the $D-$ term of
$U(1)_A$. The FI term would appear to break supersymmetry at the string
scale, but certain scalar fields are triggered to acquire large VEV's
along $D-$ and $F-$ flat directions, such that the new ``restabilized''
vacuum is supersymmetric. The complete set of $D-$ and $F-$ flat
directions involving the non-Abelian singlet fields for Model 5 was
classified in
\cite{cceel2}.   
%The scalar potential is minimized along
%directions in field space in which a subset of fields acquire string-scale
%VEV's such that the $D$ and $F$ term contributions are zero (the "flat
%directions"). Therefore, the first stage of the analysis is to determine
%the set of flat directions, which are the possible supersymmetric
%"restabilized vacua" for a given model.

In a given flat direction, the rank of the gauge group is
reduced, and effective mass terms and trilinear couplings
may be generated from higher order terms in the superpotential:
\begin{eqnarray}
W_M&=&\frac{\alpha_{K+2}}{M_{Pl}^{K-1}}\Psi_i \Psi_j \langle \Phi^K
\rangle \\
W_3&=&\frac{\alpha_{K+3}}{M_{Pl}^K}\Psi_i \Psi_j \Psi_k \langle \Phi^K
\rangle,
\end{eqnarray}
in which the fields which are in the flat direction are denoted by
$\Phi$, and those which are not by $\{\Psi_i\}$.
Hence, some fields acquire superheavy masses and decouple.  The
effective Yukawa couplings of the remaining light fields are typically
suppressed~\footnote{However, in the prototype model considered the
effective trilinear couplings arising from fourth order terms are
comparable in strength to the original Yukawas.} compared with 
Yukawa couplings of the original superpotential (the $\alpha_K$
coefficients are in principle calculable; for details, see \cite{cew}).

This procedure has been carried out for the prototype model in
\cite{cceel2,cceelw1}. We carry out the analysis of the implications of
the model for the flat directions that break the maximal number of
$U(1)$'s, leaving $U(1)_Y$ and $U(1)^{'}$ unbroken. The list of matter
superfields and their $U(1)_Y$ and $U(1)'$ charges are presented in Table
1. 

%In particular, $P_1'P_2'P_3'$ flat direction are presented in the
%following section as an example.  

%The gauge group of this effective
%theory includes an additional $U(1)'$ as well as hypercharge; for the sake
%of completeness, we present the list of matter superfields and their
%$U(1)_Y$ and $U(1)'$ charges in Table 1.

\section{Example: Low Energy Implications of a Representative Flat
Direction}

We choose to present the analysis of the model along a particular flat
direction~\footnote{Other flat directions
involve other interesting features, such as fermionic textures,
baryon number violation, and the possibility of intermediate scale
$U(1)'$ breaking.}. 
The flat direction we consider is
the $P_1'P_2'P_3'$ direction (in the notation of
\cite{cceel2,cceelw1,cceelw2}),
which involves the set of fields
$\{\varphi_{2},\varphi_{5},\varphi_{10},\varphi_{13},
\varphi_{27},\varphi_{29},\varphi_{30}\}$. 
%The VEV's of these fields are ...??? 

Along this flat direction, the effective mass terms which involve the
observable sector fields and the non-Abelian singlets~\footnote{We refer
the reader to \cite{cceelw1,cceelw2} for further details of the model,
such as the couplings involving the hidden sector fields.} are given by 
\begin{eqnarray}
\label{p123massw}
W_{M}&=&gh_f\bar{h}_b \langle \varphi_{27}\rangle+gh_g\bar{h}_d \langle
\varphi_{29}\rangle+
\frac{\alpha^{(1)}_{4}}{M_{Pl}}h_b\bar{h}_b \langle
\varphi_{5}\varphi_{10}\rangle +
\frac{\alpha^{(2)}_{4}}{M_{Pl}}h_b\bar{h}_b \langle
\varphi_{2}\varphi_{13}\rangle
\nonumber\\
&+&{g\over {\sqrt{2}}}(e^c_de_b+e^c_ge_a)\langle
\varphi_{30}\rangle +
{g\over {\sqrt{2}}}(\varphi_1 \varphi_{15}+\varphi_{4}\varphi_{9})\langle
\varphi_{10}\rangle +
{g\over {\sqrt{2}}}(\varphi_7 \varphi_{16}+\varphi_{9}\varphi_{12})\langle
\varphi_{2}\rangle\nonumber\\ &+&
{g\over {\sqrt{2}}}(\varphi_6
\varphi_{26}+\varphi_{8}\varphi_{23}+\varphi_{14}\varphi_{17})\langle
\varphi_{29}\rangle+\frac{\alpha^{(3)}_{4}}{M_{Pl}}\varphi_{21}\varphi_{25}
\langle \varphi_{27}\varphi_{29}\rangle
%\nonumber\\ &+&
%{g\over {\sqrt{2}}}(\bar{F}_1 F_1 +\bar{F}_2 F_2)\langle
%\varphi_{30}\rangle
%+{g\over {\sqrt{2}}}S_3S_5 \langle \varphi_{5}\rangle+
%{g\over {\sqrt{2}}}S_1S_5 \langle
%\varphi_{13}\rangle 
\, .
\end{eqnarray}
The  effective trilinear couplings involving
all fields which couple directly to the observable sector fields are
given by:
\begin{eqnarray}
\label{efftril}
W_{3}&=&gQ_cu^c_c\bar{h}_c+gQ_cd^c_bh_c
+\frac{\alpha^{(4)}_{4}}{M_{Pl}}Q_cd_d^ch_a\langle \varphi_{29} \rangle
+{g\over {\sqrt{2}}}e^c_ah_ah_c+
{g\over {\sqrt{2}}}e^c_fh_dh_c  \nonumber\\&+&
\frac{\alpha^{(1)}_{5}}{M^2_{Pl}}e^c_hh_eh_a\langle
\varphi_{5}\varphi_{27}\rangle+
\frac{\alpha^{(2)}_{5}}{M^2_{Pl}}e^c_eh_eh_a\langle
\varphi_{13}\varphi_{27}\rangle+
%gh_f\bar{h}_c\varphi_{22}+ 
gh_{b'}\bar{h}_c\varphi_{20} \; .
% \nonumber \\
%&+& \frac{\alpha^{(5)}_{4}}{M_{Pl}}S_2 S_6 \varphi_{20} \langle
%\varphi_{27}\rangle + \frac{\alpha^{(3)}_{5}}{M^2_{Pl}}S_2 S_6
%\varphi_{22} \langle \varphi_{2} \varphi_{13}\rangle +
%\frac{\alpha^{(4)}_{5}} {M^2_{Pl}}S_2 S_6 \varphi_{22} \langle
%\varphi_{5}\varphi_{10}\rangle
\end{eqnarray}

In the observable sector, the fields which remain light include both the
usual  MSSM states and exotic states such as a fourth ($SU(2)_L$ singlet)
down-type quark, extra fields with the same quantum numbers as the lepton
singlets, and extra Higgs doublets. There are other
massless states with exotic quantum numbers (including fractional electric
charge) that also remain in the low energy theory.
The $U(1)'$ charges of the light fields are family
nonuniversal (and hence is problematic with respect to FCNC).

There are some generic features of the superpotential which
are independent of the details of the soft supersymmetry breaking
parameters.  In addition to a large top-quark Yuakwa coupling ($\sim {\cal
O} (1)$) which is necessary for radiative electroweak symmetry breaking,
the couplings indicate $t-b$ and (unphysical) $\tau -\mu$ Yukawa
unification, with the
identification of the fields  $\bar{h}_c$, $h_c$ as the standard
electroweak Higgs doublets. 
There is no  elementary or effective canonical $\mu$-term involving
$\bar{h}_c$ and $h_c$, but rather non-canonical effective $\mu$ terms
involving additional Higgs doublets.
Finally, there is also a possibility of lepton- number
violating couplings; thus this model violates $R$- parity, and has no
stable LSP. 

With the knowledge of the massless spectrum at the string scale, the gauge
coupling beta-functions can be determined, and the gauge couplings can
then be run from the string scale (where they are predicted to unify) to
the electroweak scale.  We determine the
gauge coupling constant $g=0.80$ at the string scale by assuming
$\alpha_{s}=0.12$ (the experimental value) at the electroweak scale, and
evolving $g_{3}$ to the string scale.
%We find that $g=0.80$ is slightly higher than that of the MSSM, due to the
%presence of one additional vectorlike exotic quark pair. 
We then use this value as an input to determine the electroweak scale
values of the other gauge couplings by their (1-loop) renormalization
group equations (RGE's).
The low energy values of the gauge couplings are not
correct due to the exotic matter and non-standard $k_{Y}=11/3$ for this
model; however, it is surprising that $\sin^{2}{\theta_{W}} \sim 0.16$ and
$g_{2}=0.48$ are not too different from the experimental  values $0.23$ 
and $0.65$, respectively.

The string-scale values of the Yukawa couplings of
(\ref{efftril}) are calculable (with the knowledge of the VEV's
of the singlet fields in the flat direction). Utilizing the RGE's, we can
also determine the low energy values of the Yukawa coupling constants. The
running of the gauge couplings and the Yukawa couplings are shown in
Figure 1. 

\begin{figure}
\vskip -0.3truein
\centerline{
\hbox{
\epsfxsize=2.8truein
\epsfbox[70 32 545 740]{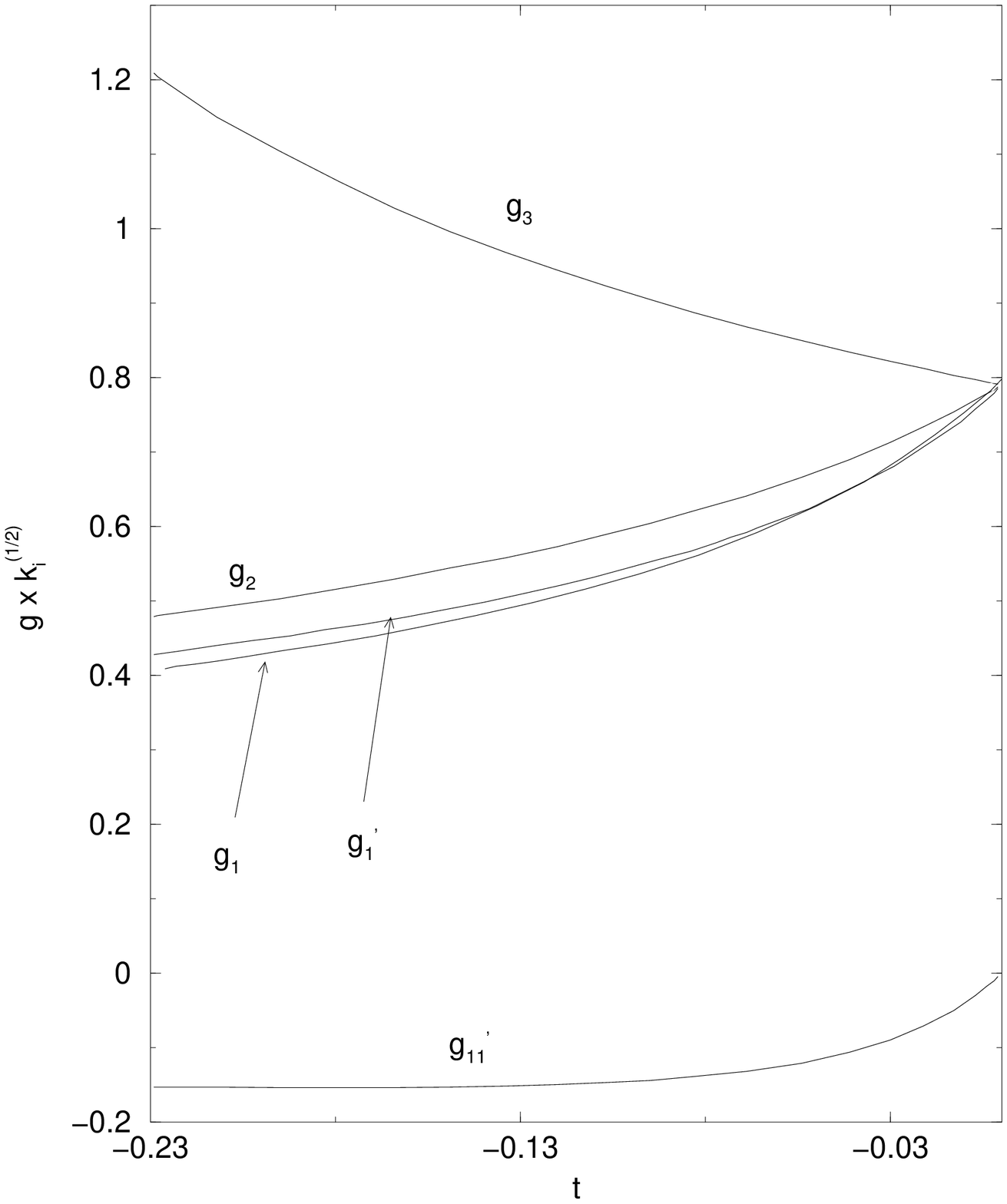}
\hskip 0.25truein
\epsfxsize=2.8truein
\epsfbox[70 32 545 740]{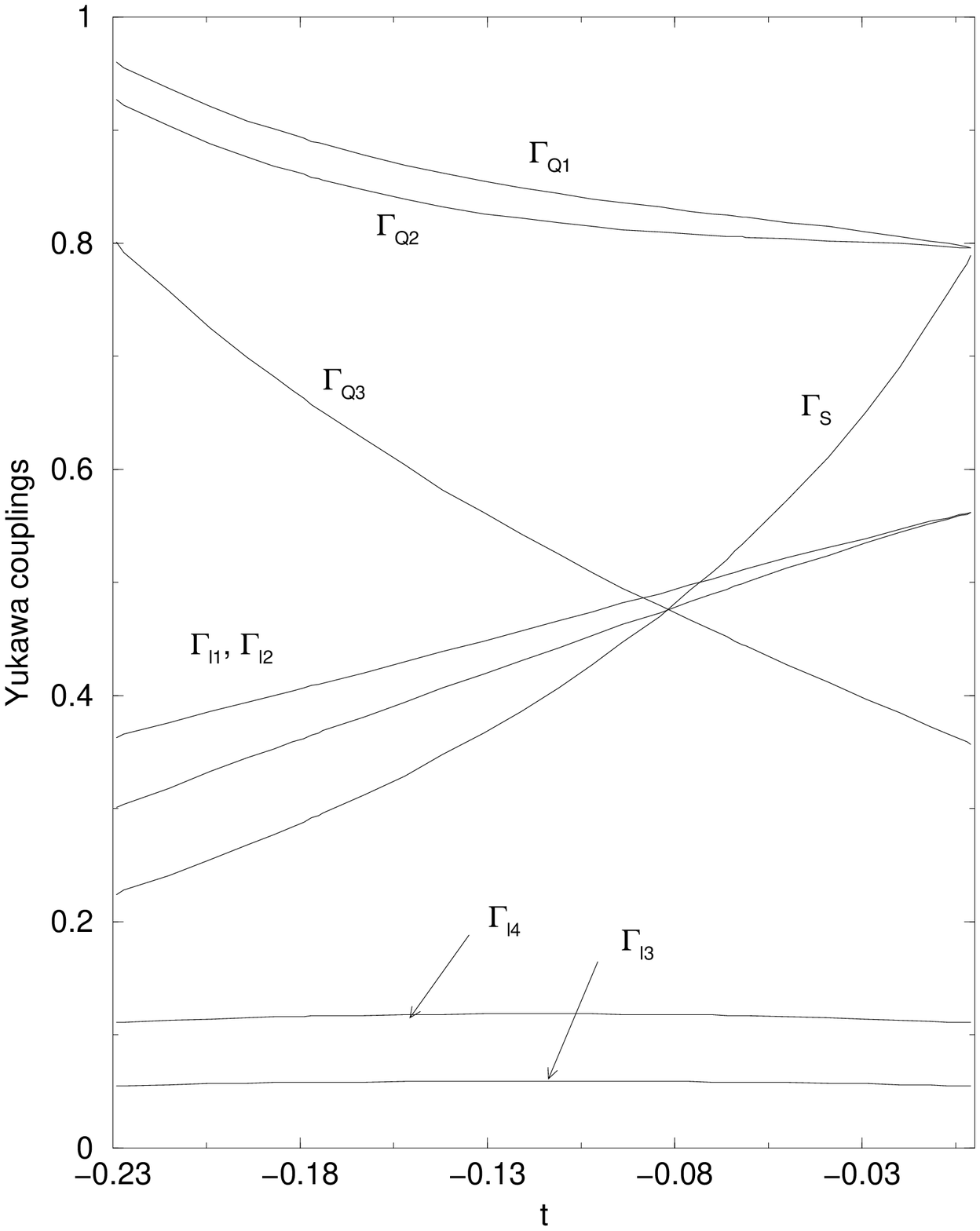}
}
}
\caption{(a): Scale variation of the gauge couplings $\times
\sqrt{k}$, with $t=(1/16\pi^2)\ln(\mu/M_{String})$, $M_{String}=5 \times
10^{17}$ GeV, and $g(M_{String})=0.80$. (b): The running of the Yukawa
couplings, in which the terms in  (\ref{efftril}) are denoted by
$\Gamma_{Q_{1,2,3}}$, $\Gamma_{l_{1,2,3,4}}$, and $\Gamma_S$,
respectively.
}
\end{figure}

%\begin{figure}
%\centerline{
%\hbox{
%\epsfxsize = 3.5truein
%\epsfbox[70 32 545 740]{hP1P2P3.1105.ps}
%}
%}
%\caption{
%Running of the Yukawa couplings for the $P_{1}^{'}P_2^{'}P_3^{'}$ direction.
% The two free parameters $\lambda_{1}$ and $\lambda_{2}$
%are chosen to be $0.9$ and $0.4$, respectively.
%}

%\end{figure}

To address the gauge symmetry breaking scenarios for this model, we
introduce soft supersymmetry breaking mass parameters
and run the RGE's from the string scale to
the electroweak scale. While the qualitative features of the analysis are
independent of the details of the soft breaking, we choose to illustrate
the analysis with a specific example with a realistic $Z-Z'$ hierarchy.    
General considerations \cite{cl,cdeel,cceel1,lw} and an inspection of
the $U(1)'$ charges of the light fields indicate that, in this example, 
the $U(1)'$ breaking is at the electroweak (TeV) scale.  Due to the lack
of a canonical effective $\mu$ term between $\bar{h}_c$, $h_c$ and a
singlet, an extended Higgs sector is required, with an
additional Higgs doublet and singlet ($\bar{h}_c$, $h_c$, $h_{b'}$, and $s
\equiv \varphi_{20}$). The symmetry breaking is characterized by a large 
(${\cal O}({\rm TeV})$) value of the SM singlet VEV, with the
electroweak symmetry breaking at a lower scale due to accidental
cancellations.  

We now present the mass spectrum for a concrete numerical example of this
scenario, which requires mild tuning of the soft supersymmetry breaking
mass parameters at the string scale. The initial and final values of the
parameters for this example are listed in Table 2.

\begin{itemize}

\item {\bf Fermion Masses:}
The masses for the $t$, $b$, $\tau$, and $\mu$ are due to Yukawa
couplings of the original superpotential, as shown in (\ref{efftril}). 
With the identification of $Q_c$ as the quark doublet of the third
family and $h_d$, $h_a$ as the lepton doublets of the third and second
families, respectively, 
$m_t=156$ GeV, $m_{b}=83$ GeV, $m_{\tau}=32$ GeV, and $m_{\mu}=27$ GeV. 
 The ratio $m_b/m_\tau$ is larger than in the usual $b-\tau$ unification
 because of the ratio $1:1/\sqrt{2}$ of the Yukawa couplings at the string
scale, and is probably inconsistent with experiment \cite{mbmtau} (of
course, the high values for $m_b$, $m_{\tau}$, and $m_\mu$ are
unphysical). Finally,  $u$, $d$, $c$, $s$, and $e^-$ remain massless.

\item {\bf Squarks/Sleptons:} 
To ensure a large $M_{Z'}$ in this model, the
squark and slepton masses have values in the
several TeV range, with $m_{\tilde{t}\,L}=2540$ GeV,
$m_{\tilde{t}\,R}=2900$
GeV; $m_{\tilde{b}\,L}=2600$ GeV, $m_{\tilde{b}\,R}=2780$ GeV;
$m_{\tilde{\tau}\,L}=2760$ GeV, $m_{\tilde{\tau}\,R}=3650$ GeV;
$m_{\tilde{\mu}\,L}=2790$ GeV, and $m_{\tilde{\mu}\,R}=3670$ GeV.

\item {\bf Charginos/Neutralinos:} 
The positively charged gauginos and higgsinos are
$\tilde{W}^{+}$, $\tilde{\bar{h}}_c$, $\tilde{\bar{h}}_a $,
and the negatively charged  gauginos and higgsinos are $\tilde{W}^{-}$,
$\tilde{h}_c$, $\tilde{h}_b'$.   There is one massless chargino, and the
other two have masses
$m_{\tilde{\chi}_{1}^{\pm}}= 591$ GeV, and
$m_{\tilde{\chi}_{2}^{\pm}}= 826$ GeV.
%(corresponding to an initial gaugino mass of 1695 GeV).

The neutralino sector consists of $\tilde{B}'$, $\tilde{B}$,
 $\tilde{W_3}$, $\tilde{\bar{h}}_{c}^0$, $\tilde{h}_{c}^0$,
$\tilde{h}_{b}^{'0}$,
 $\tilde{\varphi}^{'}_{20}$, and $\tilde{\bar{h}}_{a}^0$.
The mass eigenvalues are 
$m_{\tilde{\chi}_{1}}^0= 963$ GeV,
$m_{\tilde{\chi}_{2}}^0= 825$ GeV,
$m_{\tilde{\chi}_{3}}^0= 801$ GeV,
$m_{\tilde{\chi}_{4}}^0= 592$ GeV,
$m_{\tilde{\chi}_{5}}^0= 562$ GeV,
$m_{\tilde{\chi}_{6}}^0= 440$ GeV,
$m_{\tilde{\chi}_{7}}^0= 2$ GeV,
 and $m_{\tilde{\chi}_{8}}^0= 0$.

In both cases, the absence of couplings of the Higgs field $\bar{h}_a$
in the superpotential leads to a massless chargino and neutralino state.
The absence of an effective $\mu$ term involving $h_c$ leads to an
additional global $U(1)$ symmetry in the scalar potential, and an
ultralight neutralino pair in the mass spectrum.

\item {\bf Exotics:}
There are a number of exotic states, including
the $SU(2)_L$ singlet down-type quark, four $SU(2)_L$ singlets with
unit charge (the $e$ and extra $e^c$ states), and a number of
SM singlet ($\varphi$) states, as well as exotics
associated with the hidden sector. The scalar components of these
exotics are expected to acquire TeV-scale masses by soft supersymmetry
breaking. However, there is no mechanism within our assumptions to give
the fermions significant masses.

\item {\bf Higgs Sector:}
The non-minimal Higgs sector of three complex doublets and one
complex singlet leads to additional Higgs
bosons compared to the MSSM.  Four of the
fourteen degrees of freedom are  eaten to become the longitudinal
components of the $W^{\pm}$, $Z$, and $Z'$; and the global 
 $U(1)$ symmetry is broken, leading to a massless Goldstone boson (which,
however, acquires a small mass at the loop level) in the spectrum. 

The spectrum of the physical Higgs bosons after symmetry breaking
consists of two pairs of charged Higgs bosons $H_{1,2}^{\pm}$, four
neutral CP even Higgs scalars $(h^0_i\, ,i=1,2,3,4)$, and one CP odd Higgs
$A^0$, with masses  
$m_{H^{\pm}_1}=10$ GeV, 
$m_{H^{\pm}_2}=1650$ GeV, 
$m_{h^0_{1}}=33$
GeV, $m_{h^0_{2}}=47$ GeV, $m_{h^0_{3}}=736$ GeV, $m_{h^0_{4}}=1650$ GeV,
and $m_{A^0}=1650$ GeV.

In this model, the bound on the lightest Higgs scalar is
different that the traditional bound in the MSSM.  It is
associated with the breaking scale of the additional global $U(1)$
symmetry; since this scale is comparable to the electroweak
scale, not only one but two Higgs scalars will be
light in the decoupling limit.  In particular, the lightest Higgs
mass satisfies the (tree-level) bound~\cite{comelli}
\begin{eqnarray}
\label{bound1}
m^2_{h^0_{1}}\leq \frac{G^2}{4}v_1^2+g^2_{1'}Q_{1}^2v_1^2=(35 \, {\rm
GeV})^2.
\end{eqnarray}
%It is also possible to place a bound on the second-lightest neutral Higgs
%scalar~\cite{comelli}
%\begin{equation}
%\begin{array}{ccc}
%m^2_{h^0_{2}} & \leq &
%m^2_{h^0_{1}}+\frac{v^2}{v_2^2+v_3^2}[( \frac{G^{2}v_1^2}{4}+
%g^2_{1'}Q_{1}^2v_1^2-m^2_{h^0_{1}}) ^{\frac{1}{2}}  \\
% & + &
%(\frac{G^2}{4v^2}(v_2^2-v_1^2-v_3^2 ) ^2+
%\frac{g^2_{1'}}{v^2}(Q_1v_1^2+Q_2v_2^2+Q_3v_3^2)^2     \\
% & + & 2\frac{\Gamma_s^2}{v^2}v_2^2v_3^2
%-m^2_{h^0_{1}}) ^{\frac{1}{2}}] ^2= (85 \, {\rm GeV})^2.
%\end{array}
%\end{equation}

\end{itemize}

\section{Conclusions}

The purpose of this work has been to explore the general features
of this class of quasi-realistic superstring models through a systematic,
``top-down" analysis of a prototype model.  The results of the
investigation of the low energy implications of the mass spectrum and
couplings predicted in a subset of the restabilized vacua of this model
demonstrate that in general, the TeV scale physics is more complicated
than that of the MSSM.  

In particular, we have found that noncanonical couplings, such as mixed
effective $\mu$ terms and $R$- parity violating operators are typically
present in the superpotential. In some other cases, there are
possibilities
for potentially interesting fermion textures.  The model is also
characterized by the presence of extra matter in the low energy theory
such as SM exotics, extra $Z'$ gauge bosons with TeV scale masses, and
additional charginos, neutralinos, and Higgs bosons with patterns of
masses that differ substantially from the MSSM.

The particular model we studied is not realistic, in part due to the
presence of (ultralight or massless) extra matter.
However,  due to the large number of possible models that can be derived
from string theory, this result does not invalidate the potential
viability of string models, or the motivation for investigating their
phenomenological implications~\footnote{Progress
has been made in exploring models in which the exotic matter decouples
above the electroweak scale; e.g. see \cite{faraggi98}.}. We stress that
the features of this model are likely to be generic to this class of
quasi-realistic models based on weakly coupled heterotic string theory,
and thus warrant further consideration.

\acknowledgments
This work
was supported in part by U.S. Department of Energy Grant No. 
DOE-EY-76-02-3071.
\newpage

%========================================================================
%          MACROS FOR REFERENCES
%========================================================================
\def\B#1#2#3{\/ {\bf B#1} (19#2) #3}
\def\NPB#1#2#3{{\it Nucl.\ Phys.}\/ {\bf B#1} (19#2) #3}
\def\PLB#1#2#3{{\it Phys.\ Lett.}\/ {\bf B#1} (19#2) #3}   
\def\PRD#1#2#3{{\it Phys.\ Rev.}\/ {\bf D#1} (19#2) #3}
\def\PRL#1#2#3{{\it Phys.\ Rev.\ Lett.}\/ {\bf #1} (19#2) #3}
\def\PRT#1#2#3{{\it Phys.\ Rep.}\/ {\bf#1} (19#2) #3}
\def\MODA#1#2#3{{\it Mod.\ Phys.\ Lett.}\/ {\bf A#1} (19#2) #3}
\def\IJMP#1#2#3{{\it Int.\ J.\ Mod.\ Phys.}\/ {\bf A#1} (19#2) #3}
\def\nuvc#1#2#3{{\it Nuovo Cimento}\/ {\bf #1A} (#2) #3}
\def\RPP#1#2#3{{\it Rept.\ Prog.\ Phys.}\/ {\bf #1} (19#2) #3}
\def\etal{{\it et al\/}}
\bibliographystyle{unsrt}

\newpage

\begin{center}
\begin{tabular}{|c|c|c|c|}
\hline\hline
$(SU(3)_C,SU(2)_L,$& &
    $6Q_{Y}$&$100Q_{Y'}$\\
$SU(4)_2,SU(2)_2)$ &&      &\\
%       &     &&&&&&&&&\\
\hline\hline
(3,2,1,1):&$Q_a$& 1&68\\
&$Q_b$ &  1&68\\
&$Q_c$ & 1&$-$71\\
\hline
($\bar{3}$,1,1,1):&$u^c_a$&  $-$4&6\\
&$u^c_b$ & $-$4&6\\   
&$u^c_c$ & $-$4&$-$133\\  
&$d^c_a$ & 2&$-$3\\
&$d^c_b$ & 2&136\\
&$d^c_c$ & 2&$-$3\\
&$d^c_d$ & 2&$-$3\\
\hline
(1,2,1,1):&$\bar{h}_a$ &  3&$-$74\\
&$\bar{h}_b$ & 3&65\\
&$\bar{h}_c$ & 3&204\\
&$\bar{h}_d$ &  3&65\\
&$h_a$ &  $-$3&74\\
&$h_b$ & $-$3&$-$65\\
&$h_c$ &  $-$3&$-$65\\
&$h_d$ & $-$3&$-$65\\
&$h_e$ & $-$3&$-$204\\
&$h_f$ & $-$3&$-$65\\
&$h_g$ & $-$3&$-$65\\
\hline
(3,1,1,1):&${\cal D}_a$& $-$2&$-$136\\
\hline
\hline
\end{tabular}
\end{center}
\noindent Table Ia: List of non-Abelian non-singlet observable sector
fields in the model with their charges under hypercharge and $U(1)'$.

%\newpage
\begin{center}
\begin{tabular}{|c|c|c||c|c|c|}
\hline\hline 
 &  $6Q_{Y}$&$100Q_{Y'}$ & & $6Q_{Y}$ &$100Q_{Y'}$ \\
 &          &            & &          & \\
%       &     &&&&&&&&&\\
\hline\hline
$e^c_{a,c}$ &6&$-$9 & $e^c_b$  &6&$-$9\\
$e^c_{d,g}$ &6&130 & $e^c_e$ &6&130\\
$e^c_f$ &6&130 & $e^c_h$ &6&130\\
$e^c_i$ &6&$-$9 & $e_{a,b}$ &6&$-$130\\
$e_c$ &6&$-$130 & $e_{d,e}$ &$-$6&9\\
$e_f$ &$-$6&$-$269 & & & \\
\hline
\hline
 &  $6Q_{Y}$&$100Q_{Y'}$ & & $6Q_{Y}$ &$100Q_{Y'}$ \\
 &          &            & &          & \\
$\varphi_{1}$ &0&0 & $\varphi_{2,3}$ &0&0 \\
$\varphi_{4,5}$ &0&0 & $\varphi_{6,7}$ &0&0 \\
$\varphi_{8,9}$ &0&0 & $\varphi_{10,11}$&0&0 \\
$\varphi_{12,13}$&0&0 & $\varphi_{14,15}$&0&0\\
$\varphi_{16}$ &0&0 & $\varphi_{17}$ &0&0 \\
$\varphi_{18,19}$&0&$-$139 & $\varphi_{20,21}$&0&$-$139 \\
$\varphi_{22}$ &0&$-$139 & $\varphi_{23}$ &0&0 \\
$\varphi_{24}$ &0&0 & $\varphi_{25}$ &0&139 \\
$\varphi_{26}$ &0&0 & $\varphi_{27}$ &0&0 \\
$\varphi_{28,29}$&0&0 & $\varphi_{30}$ &0&0 \\
\hline
\hline
\end{tabular}
\end{center}
\noindent Table Ib: List of non-Abelian singlet fields in
the model with their charges under hypercharge and $U(1)'$.

\newpage

\begin{center}
\begin{tabular}{||c||c|c||c||c|c||}
\hline
\hline

 &  $M_Z$ & $M_{String}$ & & $M_Z$ & $M_{String}$  \\ \hline
$g_1$& 0.41& 0.80&$M_1$& 444&1695 \\
$g_2$& 0.48& 0.80&$M_2$& 619& 1695\\
$g_3$& 1.23& 0.80&$M_3$& 4040& 1695\\
$g_1'$& 0.43& 0.80&$M_1'$& 392& 1695\\
$\Gamma_{Q1}$&0.96 & 0.80&$A_{Q1}$ & 3664&8682\\
$\Gamma_{Q2}$&0.93 & 0.80& $A_{Q2}$&4070&9000\\
$\gamma_{Q3}$&0.27 & 0.08& $A_{Q3}$&5018&1837\\
$\Gamma_{l1}$& 0.30& 0.56&$A_{l1}$ &$-$946&4703\\
$\Gamma_{l2}$& 0.36& 0.56&$A_{l2}$ &$-$707&4532\\
$\Gamma_{l3}$& 0.06& 0.05&$A_{l3}$ &4613&4425\\
$\Gamma_{l4}$& 0.11& 0.13&$A_{l4}$ &4590&4481\\
$\Gamma_{s}$& 0.22& 0.80 &$A$      &1695&12544\\
%$\Gamma_{H1/H2}$& 0.006& 0.005& $A_{H1/H2}$ &2518&4237\\
%$\Gamma_{H3/H4}$& 0.001& 0.004& $A_{H3/H4}$ &4896&4237\\
$m_{Q_c}^2$ & $(2706)^2$ & $(2450)^2$ &
$m_{d_d}^2$ & $(4693)^2$ & $(2125)^2$ \\
$m_{u_c}^2$ & $(2649)^2$ & $(2418)^2$ &
$m_{d_c}^2$ & $(2734)^2$ & $(2486)^2$\\
$m_{\bar{h}_c}^2$ & $(1008)^2$ & $(5622)^2$ &
$m_{h_b'}^2$&$(826)^2$&$(2595)^2$\\
$m_{\varphi_{20'}}^2$ & $-(518)^2$ & $(6890)^2$ &
$m_{\varphi_{22'}}^2$ &$(3031)^2$&$(11540)^2$\\ 
$m_{h_a}^2$ & $(3626)^2$ & $(3982)^2$ &
$m_{h_c}^2$ & $-(224)^2$&$(5633)^2$\\
$m_{h_d}^2$ & $(3666)^2$ & $(4100)^2$ &
$m_{h_e}^2$ & $(4274)^2$ & $(4246)^2$\\ 
$m_{e_a}^2$ & $(2770)^2$ & $(3564)^2$ &
$m_{e_f}^2$ & $(2780)^2$ & $(3958)^2$\\
$m_{e_e}^2$ & $(4195)^2$ & $(4254)^2$ &
$m_{e_h}^2$ & $(4259)^2$ & $(4236)^2$\\
%$m_{H_3}^2$ & $(4327)^2$ & $(4237)^2$ & $m_{H_4}^2$
%&$(4327)^2$&$(4237)^2$\\
%$m_{H_5}^2$& $(4315)^2$ & $(4237)^2$ & $m_{H_7}^2$
%&$(4315)^2$&$(4237)^2$ \\
\hline\hline
\end{tabular}
\end{center}
\noindent Table II: $P_1'P_2'P_3'$ flat direction: values of the parameters at 
$M_{String}$ and $M_{Z}$, with $M_{Z'}=735$
GeV and $\alpha_{Z-Z'}=0.005$.  All mass parameters are given in GeV. 

%\end{table}

\end{document}